\documentclass[]{article}
\usepackage{graphicx}
\usepackage{amsmath}
\usepackage{amsfonts}
\usepackage{cite}
\usepackage[dvipdfm,colorlinks,linkcolor=blue,citecolor=blue,urlcolor=blue,anchorcolor=blue]{hyperref}
\usepackage{geometry}
\usepackage{CJK}

\begin{document}

\title{Controllable Multi-Wave Mixing Talbot Effect}

\author{Yiqi Zhang, Xin Yao, Chenzhi Yuan, Peiying Li, Jiamin Yuan, \\
       Weikang Feng, Shuqiao Jia, and Yanpeng Zhang$^*$ \\
       \textit{Key Laboratory for Physical Electronics and Devices of the Ministry of Education \&} \\
       \textit{Shaanxi Key Lab of Information Photonic Technique,} \\
       \textit{Xi'an Jiaotong University, Xi'an 710049, China} \\
       $^*$corresponding author: ypzhang@mail.xjtu.edu.cn}
\date{}

\maketitle

\maketitle

\begin{abstract}
We theoretically study the Talbot effects resulted from the four-wave mixing and six-wave mixing signals,
which are periodically modulated due to the coherence control effect.
Corresponding to different dressing states,
the enhancement and suppression conditions,
that will affect the properties of the multi-wave mixing signals are also discussed in detail.
Such proposal can be useful in all optical-controlled pattern formation and propagation of light.
\end{abstract}

\section{Introduction}

Talbot effect, which was observed by H. F. Talbot in 1836 \cite{talbot_1836}
and first analytically explained by Lord Rayleigh in 1881 \cite{rayleigh_1881},
is a near-field diffraction phenomenon
that the light field spatially imprinted with periodic structure can have self-imaging
at certain periodic imaging planes (the so-called Talbot planes).
Such self-imaging effect holds a range of applications from image preprocessing and synthesis,
photolithography, optical testing, optical metrology, spectrometry, to optical computing \cite{patorski_1989}.

To date, studies on Talbot effect have been reported associated with
atomic waves \cite{chapman_pra_1995,ryu_prl_2006},
surface waves \cite{dennis_oe_2007,maradudin_njp_2009},
nonclassical light \cite{muller_pra_2009},
or waveguide arrays \cite{iwanow_prl_2005}.
Among the quiverful of researches,
the nonlinear Talbot effect with second-harmonic (SH)
were demonstrated both experimentally \cite{zhang_prl_2010}
and theoretically \cite{wen_josab_211}
in periodically poled LiTaO$_3$ (PPLT) crystal for the first time.
The Talbot effect with beam imprinted with an electromagnetically induced grating (EIG) has been also proposed \cite{wen_apl_2011}.
However, no progress in higher nonlinear optical process was reported before as far as to our knowledge.

The crucial point and the guiding ideology to observe the Talbot effect
is how to produce a spatial periodical incidence in transverse dimension.
From this point of view,
it is natural to realize that researchers have created
spatially periodic four-wave mixing (FWM) and six-wave mixing (SWM) signals
due to the periodic atomic coherence induced by standing waves (SWs)
\cite{andre_prl_2002,artoni_prl_2006,gao_ol_2010,wu_josab_2008},
which can play an important role in lasing without inversion \cite{imamoglu_ol_1989},
slow light generation \cite{wang_oe_2012}, photon controlling and information
storage \cite{lukin_nature_2001, phillips_prl_2001, liu_nature_2001}, fiber lasers \cite{liu_oe_2005, liu_pra_2008}, and
so on.
The periodic pattern of the multi-wave mixing (MWM) signals can be flexibly controllable
by adjusting the atomic coherence via changing the beam detunings.

In this article, we investigate the MWM Talbot effect for the first time.
And as the coexisting FWM and SWM processes have been observed in the Cd (S, Se) semiconductor doped glasses \cite{ma_prl_1993},
the idea can be also executed in some solid crystals such as Pr-doped YSO crystals \cite{ma_prl_1993,ham_oc_1997,klein_prl_1999,wang_apl_2008} besides atomic vapors.
The article is organized as follows:
in Sec. \ref{theory}, we introduce the basic theory includes the energy level systems,
some conceptions, and how to prepare the periodic FWM and MWM signals;
in Sec. \ref{esc}, we discuss the enhancement and suppression conditions,
due to which the energy levels split;
in Sec. \ref{talbot}, we investigate the Talbot effects of the periodic MWM signals;
and in the last section, we conclude the article.

\section{Theoretical Model and Analysis}
\label{theory}

\begin{figure}[htbp]
  \centering
  \includegraphics[width=0.8\textwidth]{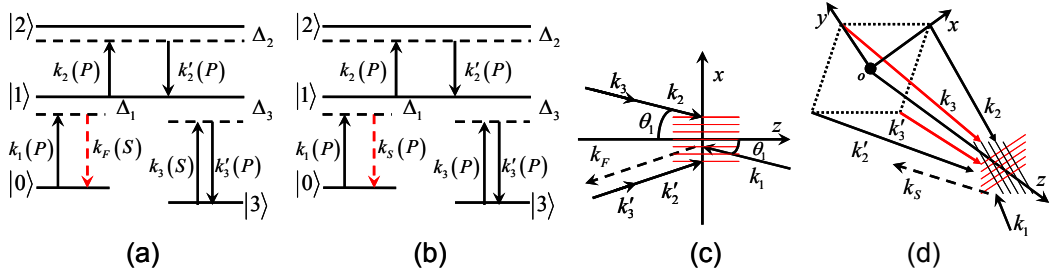}
  \caption{
  The schematic of the reverse Y-type atomic system to produce MWM signals.
  (a) ${\bf k}_F$ represents the singly- or doubly-dressed FWM corresponds to a weak or strong ${\bf k}_3$, respectively.
  (b) Emission of the doubly-dressed SWM signal ${\bf k}_S$.
  The letters $S$ and $P$ in the brackets mean the polarization of the wave.
  (c) and (d) The Cartesian geometric configurations to generate FWM and SWM signals, respectively;
  The SWs along $x$ and $y$ are constructed by ${\bf k}_2$, ${\bf k}_2'$ and ${\bf k}_3$, ${\bf k}_3'$, respectively.
  }
  \label{fig1}
\end{figure}

We consider the FWM and SWM processes in the reverse Y-type atomic levels system derived by five beams,
as shown in Figs. \ref{fig1}(a) and (b).
A candidate for the systems are rubidium atomic vapors.
Also, in Pr:YSO crystal,
the energy-level system with  $|0\rangle = |^3H_4(\pm3/2)\rangle$,
$|1\rangle = |^1D_2(\pm3/2)\rangle$,
$|3\rangle = |^3H_4(\pm1/2)\rangle$
can be used as a sub-system of our proposed system.
Here we note that to guarantee the applicability of our proposal to all the possible candidate systems,
we choose parameters in simulation arbitrarily.
Figs. \ref{fig1}(c) and (d) are the corresponding beam geometric configurations.
The transition $\left| 0 \right\rangle  \to \left| 1 \right\rangle $ with resonant frequency $\Omega_{10} $
is probed by beam ${\bf E}_1 $ (with frequency $\omega_1$, frequency detuning $\Delta_1$ and wave-vector ${\bf{k}}_1$);
$\left| 1 \right\rangle  \to \left| 2 \right\rangle $ with resonant frequency $\Omega_{21} $
is pumped by beams ${\bf E}_2$ (with $\omega_2$, $\Delta_2$ and ${\bf{k}}_2$) and
${\bf E}_2'$ (with $\omega_2$, $\Delta_2$ and ${\bf{k}}_2'$);
$\left| 3 \right\rangle  \to \left| 1 \right\rangle$ with resonant frequency $\Omega_{31} $
is pumped by beams ${\bf E}_3$ (with $\omega_3$, $\Delta_3$ and ${\bf{k}}_3$) and
${\bf E}_3'$ (with $\omega_3$, $\Delta_3$ and ${\bf{k}}_3'$).
The frequency detunings are defined as $\Delta_1  = \Omega_{10}  - \omega_1$,
$\Delta_2  = \Omega_{21}  - \omega_2$, and $\Delta_3  = \Omega_{13}  - \omega_3 $.
In the Cartesian coordinate introduced in Figs. \ref{fig1}(c) and (d),
the wave-vectors are elaborated designed that ${\bf E}_1$ propagates  $z$-negative with an angle $\theta_1$;
${\bf E}_2$ and ${\bf E}_3$ propagate along the opposite direction of ${\bf{k}}_1$,
deviating from the $z$-positive direction with $\theta_1$;
${\bf E}_2'$ and ${\bf E}_3'$ propagate symmetrically to ${\bf E}_2$ and ${\bf E}_3$
with respect to  $z$-negative direction.

As shown in Fig. \ref{fig1}(a),
if the fields ${\bf E}_2$, ${\bf E}_2'$, ${\bf E}_3$ and ${\bf E}_3'$ are not strong  sufficiently,
we consider the undressed FWM signal ${\bf E}_F$ with $\omega_F  = \omega_1$ and ${\bf{k}}_F  = {\bf{k}}_1  + {\bf{k}}_3  - {{\bf k}}_3'$;
if ${\bf E}_3$ and ${\bf E}_3'$ are weak but ${\bf E}_2$ and ${\bf E}_2'$ are strong sufficiently to induce dressing effect (act as the dressing fields),
we consider the singly-dressed FWM signal ${\bf E}_{F1}$ with $\omega_{F1}  = \omega_1$ and ${{\bf k}}_{F1} = {{\bf k}}_F$.
While if ${\bf E}_3$ and ${\bf E}_3'$ are also strong sufficiently to induce dressing effect,
we consider the doubly-dressed FWM signal ${\bf E}_{F2}$ with $\omega_{F2}  = \omega_1$ and ${{\bf k}}_{F2} = {{\bf k}}_F$.
Even though ${\bf E}_2$ and ${\bf E}_2'$ will induce another FWM signal ($P$ polarization),
we still can get the expected FWM signal via a polaroid for its polarization selectivity.
By using five beams with polarizations as shown in Fig. \ref{fig1}(b),
a SWM signal ${\bf E}_S$ with $\omega_S  = \omega_1$ and
${{\bf k}}_S  = {{\bf k}}_1  + {{\bf k}}_2  - {{\bf k}}_2'  + {{\bf k}}_3  - {{\bf k}}_3'$ can be obtained.

To obtain the MWM signal,
we solve the density matrix equations and the Liouville pathways
for $\rho_{10}^{\left( 3 \right)}$ and $\rho_{10}^{\left( 5 \right)}$,
the amplitude of which is proportional to that of the MWM signal.
First, for the undressed FWM signal ${\bf E}_{F}$,
we can obtain $\rho_{10}^{\left( 3 \right)}$ via the Liouville pathway \cite{zhang_2009} as
\begin{equation}\label{eq1}
\rho_{00}^{\left( 0 \right)} \xrightarrow{{\bf E}_1}
\rho_{10}^{\left( 1 \right)} \xrightarrow{({\bf E}_3)^*}
\rho_{30}^{\left( 2 \right)} \xrightarrow{{\bf E}_3'}
\rho_{10}^{\left( 3 \right)},
\end{equation}
\begin{equation}\label{eq2}
\rho_{10}^{\left( 3 \right)}  = \frac{{ - iG_{{\rm{FWM}}} }}{{d_1^2 d_3 }},
\end{equation}
where $G_{\rm{FWM}} = G_{10} G_{30} (G_{30}')^*  \exp(i{\bf{k}}_F  \cdot {\bf{r}})\exp(-i\omega_1 t)$,
$d_1  = \Gamma_{10}  + i\Delta_1 $, $d_3  = \Gamma_{30}  + i\left( {\Delta_1  - \Delta_3 } \right)$.
The terms $G_{i0}  = \mu_{mn} {E}_{i0} /\hbar$ and $G_{i0}'  = \mu_{mn} {E}_{i0}' /\hbar$ with $i = 1,2$
are the amplitudes of the Rabi frequencies of ${\bf E}_i$ and ${\bf E}_i'$, respectively,
where $\mu_{mn}$ is the dipole moment of transition between $\left| m \right\rangle$ and $\left| n \right\rangle$,
which is driven by ${\bf E}_i$;
${E}_{i0}$ and ${E}_{i0}'$ are the amplitudes of ${\bf E}_i$ and ${\bf E}_i'$, respectively.
$\Gamma_{pq}$ represents the transverse relaxation rate between $|p\rangle$ and $|q\rangle$ for $p \ne q$,
and the longitudinal relaxation rate of $|p\rangle$ for $p=q$.
It is obvious to see that the amplitude of $\rho_{10}^{\left( 3 \right)}$ in Eq. (\ref{eq2}) is spatial independent,
so ${\bf E}_F$ is uniform.

Next, we consider the singly-dressed FWM process from ${\bf E}_2$ and ${\bf E}_2'$,
which are strong sufficiently.
In the spatial interaction region,
${\bf E}_2$ and ${\bf E}_2'$ will interfere with each other and create a SW,
which leads to periodic Rabi frequency amplitude $\left|G_{2t} (x) \right|^2  = G_{20}^2  + G_{20}'^2  + 2G_{20} G_{20}' \cos \left[ {2\left( {k_2 \sin \theta_1 } \right)x} \right]$.
Therefore, the Liouville pathway for ${\bf E}_{F1}$ will be modified into:
\begin{equation}\label{eq3}
\rho_{00}^{(0)} \xrightarrow{{\bf E}_1}
\rho_{|G_{2t}(x) \pm 0\rangle}^{(1)} \xrightarrow{({\bf E}_3)^*}
\rho_{30}^{(2)} \xrightarrow{{\bf E}_3'}
\rho_{|G_{2t}(x) \pm 0\rangle}^{(3)},
\end{equation}
where $|G_{2t}(x) \pm \rangle$ represent two dressed states produced by the spatial periodic dressing effect.
According to Eq. (\ref{eq3}), we can obtain the spatial-dependent dressed density matrix element as:
\begin{equation}\label{eq4}
\rho_{F1}^{\left( 3 \right)} \left( x \right) = \frac{{ - iG_{{\rm{FWM}}} }}{\left[d_1  + |G_{2t}(x)|^2/d_2 \right]^2 d_3 },
\end{equation}
where $d_2  = \Gamma_{20}  + i\left( {\Delta_1  + \Delta_2 } \right)$.
The amplitude of $\rho_{F1}^{\left( 3 \right)}$ shows obvious periodic variation along $x$ direction with a period of $d_x  = \lambda_2/(2\sin \theta_1)$.

When ${\bf E}_3$ and ${\bf E}_3'$ get strong significantly,
we will get the doubly-dressed FWM signal ${\bf E}_{F2}$.
The Liouville pathway and the doubly-dressed density matrix element for this case are
\begin{equation}\label{eq5}
\rho_{00}^{(0)} \xrightarrow{{\bf E}_1}
\rho_{|G_{3t}\pm G_{2t}(x) \pm 0\rangle}^{(1)} \xrightarrow{({\bf E}_3)^*}
\rho_{30}^{(2)} \xrightarrow{{\bf E}_3'}
\rho_{|G_{3t}\pm G_{2t}(x) \pm 0\rangle}^{(3)} ,
\end{equation}
\begin{equation}\label{eq6}
\rho_{F2}^{(3)}(x) = \frac{- iG_{\rm{FWM}}}{\left[d_1 + |G_{30}|^2/d_3 + |G_{2t}(x)|^2/d_2 \right]^2 d_3}.
\end{equation}

In Fig. \ref{fig1}(b),
the spatially periodic dressing effect from ${\bf E}_2$ and ${\bf E}_2'$ remains,
and the dressing fields ${\bf E}_3$ and ${\bf E}_3'$ with same polarization
(compared to the case of Eq. (\ref{eq6}) with ${\bf E}_3$ and ${\bf E}_3'$ having orthogonal polarizations)
can also interfere with each other
and create periodic Rabi frequency amplitude
$|G_{3t} (y)|^2  = G_{30}^2  + G_{30}'^2  + 2G_{30} G_{30}' \cos \left[ {2\left( {k_3 \sin \theta_2 } \right)y} \right]$.
Therefore, this SWM will suffer from two dressing effects,
which are both spatial periodic with different periods $d_x$ and $d_y  = \lambda_3/(2\sin \theta_2)$, respectively.
The corresponding Liouville pathway and doubly-dressed density matrix element are
\begin{equation}\label{eq7}
\rho_{00}^{(0)} \xrightarrow{{\bf E}_1}
\rho_{|G_{3t}(y) \pm G_{2t}(x) \pm 0\rangle}^{(1)} \xrightarrow{({\bf E}_3)^*}
\rho_{30}^{(2)} \xrightarrow{{\bf E}_3'}
\rho_{|G_{3t}(y) \pm G_{2t}(x) \pm 0\rangle}^{(3)} \xrightarrow{({\bf E}_2)^*}
\rho_{2 \pm 0}^{(4)} \xrightarrow{{\bf E}_2'}
\rho_{|G_{3t}(y) \pm G_{2t}(x) \pm 0\rangle}^{(5)} ,
\end{equation}

\begin{equation}\label{eq8}
\rho_{10}^{\left( 5 \right)}  = \frac{{iG_{{\rm{SWM}}} }}{{\left[d_1  + |G_{2t}(x)|^2/d_2 + |G_{3t}(y)|^2/d_3 \right]^3 d_2 d_3 }},
\end{equation}
where $G_{\rm{SWM}} = G_{10} G_{20} (G_{20}')^*  G_{30} (G_{30}')^*\exp(i{\bf{k}}_S  \cdot {\bf{r}})\exp(-i\omega_1 t)$.

With these expressions, the spatial periodic pattern formation of MWM signals can be investigated.
It is clear that the periodical variation of singly-dressed FWM,
doubly-dressed FWM and doubly-dressed SWM signals with respect to  $x$-axis at $z=0$,
are all derived from the periodical dressing effects.
With different probe detunings, these MWM signals will show profiles with significant differences.
So, in order to analyze such dependence,
we consider the so-called enhancement and suppression conditions due to dressing effect
before investigating the MWM Talbot effects.

\section{Suppression and Enhancement Conditions}
\label{esc}

\begin{figure*}
  \centering
  \includegraphics[width=0.8\textwidth]{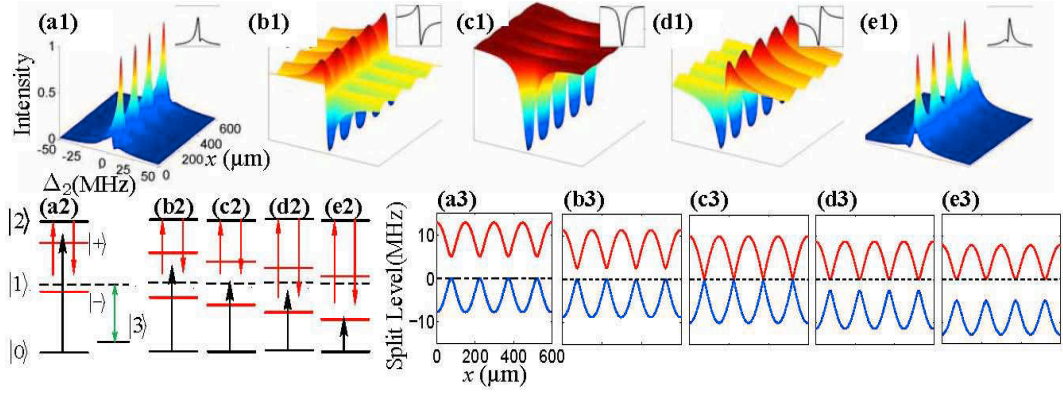}
  \caption{The normalized intensities versus $\Delta_2$ and $x$ of singly-dressed FWM with
  (a1) enhancement effect,
  (b1) enhancement-suppression effect,
  (c1) suppression effect,
  (d1) suppression-enhancement effect,
  and (e1) enhancement effect,
  corresponding to $\Delta_1=-5~\rm{MHz}$,
  $-4~\rm{MHz}$, $0$, $4~\rm{MHz}$,
  and $5~\rm{MHz}$, respectively.
  The insets are the incident intensity versus $\Delta_2$.
  (a2)-(e2) The corresponding split energy level (red lines marked $|+\rangle$ and $|-\rangle$ on the right are split energy levels,
  the dotted line is the initial energy level) diagrams with dressing fields at $x=0$.
  (a3)-(e3) The corresponding periodic split energy levels versus $x$.
  The other parameters are
  $G_2=15~\rm{MHz}$, $\Gamma_{10}=5~\rm{MHz}$, and $\Gamma_{20}=1~\rm{kHz}$.
  }
  \label{fig2}
\end{figure*}

The characteristics of FWM are affected by the dressing effect,
which depends on $\Delta_1$, $\Delta_2$,
and $\left| {G_{2t} \left( x \right)} \right|^2 $.
As shown in Fig. \ref{fig2},
because of $\left| {G_{2t} \left( x \right)} \right|^2 $,
$|1\rangle$ is split into two dressing states
$|G_{2t}(x)\pm\rangle$ with eigen-frequencies
$\lambda_{|G_{2t}(x)\pm\rangle}  = \Delta_2/2 \pm \sqrt{\Delta_2^2/4 +|G_{2t}(x)|^2}$.
Therefore, $|G_{2t}(x)\pm\rangle$ are periodic along $x$.

Absorption will be enhanced when the probe resonates with the dressing states, i.e.,
$\Delta_1  =  - \lambda_{|G_{2t}(x) \pm \rangle}$,
which corresponds to the electromagnetically induced absorption (EIA) condition.
Accordingly, the FWM signal will get enhanced resonantly.
Thus we define $\Delta_1  =  - \lambda_{|G_{2t}(x) \pm \rangle}$ as the enhancement condition.
While when the probe reaches two-photon resonance ($\Delta_1  + \Delta_2  = 0$),
absorption will be suppressed (the EIT condition),
and the FWM signal in such case will be suppressed correspondingly.
Thus, we define $\Delta_1  + \Delta_2  = 0$ as the suppression condition.
So, if $\Delta_1$ is set at discrete values orderly from negative to positive
and $\Delta_2$ is scanned, we can obtain the periodic enhancement/suppression condition
and FWM signals along $x$.
In Figs. \ref{fig2}(a1)-(e1),
we display the periodic FWM signals along $x$ corresponding to different $\Delta_1$,
and the insets are the dependence of the FWM signal intensity on $\Delta_2$ at $z=0$ and $x=0$.
In Figs. \ref{fig2}(a2)-(e2), we exhibit the corresponding split energy level diagrams,
and in Figs. \ref{fig2}(a3)-(e3) the corresponding periodic properties of the split energy levels along $x$.
In Fig. \ref{fig2}(a1), $\Delta_1$ is relatively large and positive,
and we only consider the enhancement because of the weakness of the suppression.
When $\Delta_1$ decreases but is still positive, $\Delta_2$ will meet the enhancement condition
($\Delta_1  =  - \lambda_{|G_{2t}(x)+\rangle}$) first,
and then the suppression condition as shown in Fig. \ref{fig2}(b1).
Because of $\Delta_1  = 0$ in Fig. \ref{fig2}(c1), so that $\Delta_2$ can never meet the enhancement condition.
The case in Fig. \ref{fig2}(d1) is opposite to that in Fig. \ref{fig2}(b1).
And in Fig. \ref{fig2}(e1), the enhancement is dominant again, the same as that in Fig. \ref{fig2}(a1),
for $\Delta_1$ is relatively large and negative.

For the doubly-dressed MWM signals ${\bf E}_{F2}$ and ${\bf E}_S$,
we take ${\bf E}_{F2}$ to execute our discussions.
First, the interference between ${\bf E}_3$ and ${\bf E}_3'$ leads to periodic dressing effect,
and therefore split the naked state into two dressing states $|G_{3t}(y)\pm\rangle$
with eigen-frequencies $\lambda_{|G_{3t}(y)\pm\rangle}  = - \Delta_3/2 \pm \sqrt{\Delta_3^2/4+ |G_{3t}(y)|^2}$.
${\bf E}_2$ and ${\bf E}_2'$ also act as dressing fields,
so the first-order dressing state $|G_{3t}(y)\pm\rangle$
will split into second-order dressing states $|G_{3t}(y)\pm G_{2t}(x)\pm\rangle$ with frequencies
\begin{equation}\label{eq10}
\lambda_{|G_{3t}(y) - G_{2t}(x) \pm \rangle}  =
 \frac{-\Delta_3  - \sqrt{\Delta_3^2  + 4|G_{3t}(y)|^2}}{2}
 + \frac{\Delta_2'  \pm \sqrt {\Delta_2'^2  + 4|G_{2t}(x)|^2}}{2},
\end{equation}
\begin{equation}\label{eq9}
\lambda_{|G_{3t}(y) + G_{2t}(x) \pm \rangle}  = \frac{-\Delta_3  + \sqrt{\Delta_3^2  + 4|G_{3t}(y)|^2}}{2}
+ \frac{\Delta_2''  \pm \sqrt {\Delta_2''^2  + 4|G_{2t}(x)|^2}}{2},
\end{equation}
where $\Delta_2' = \Delta_2  - \{- \Delta_3  - \sqrt{\Delta_3^2+4|G_{3t}(y)|^2}\}/2$,
$\Delta_2'' = \Delta_2  -\{- \Delta_3  + \sqrt{\Delta_3^2+4|G_{3t}(y)|^2}\}/2$, respectively.

\begin{figure*}
  \centering
  \includegraphics[width=1\textwidth]{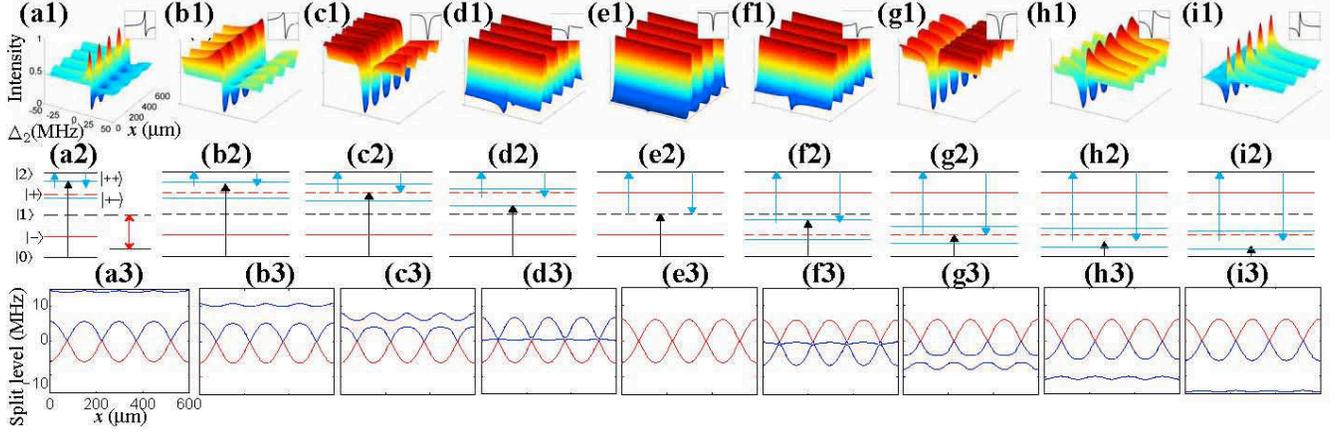}
  \caption{The normalized intensities versus $\Delta_2$ and $x$ of
  (a1) doubly-dressed FWM with enhancement effect,
  (b1) enhancement-suppression effect,
  (c1) suppression effect,
  (d1) suppression-enhancement effect,
  (e1) suppression effect,
  (f1) enhancement-suppression effect,
  (g1) suppression effect,
  (h1) suppression-enhancement effect,
  and (i1) enhancement effect,
  corresponding to $\Delta_1=-18~\rm{MHz}$,
  $-8~\rm{MHz}$, $-5~\rm{MHz}$, $-2~\rm{MHz}$, $0$, $2~\rm{MHz}$, $5~\rm{MHz}$,
  $8~\rm{MHz}$, and $18~\rm{MHz}$, respectively.
  The insets are the incident intensity versus $\Delta_2$.
  (a2)-(i2) The corresponding split energy level diagrams with dressing fields at $x=0$.
  (a3)-(i3) The corresponding periodic split energy levels versus $x$.
  The other parameters are
  $\Delta_3=0$, $G_2=G_3=15~\rm{MHz}$, $\Gamma_{10}=5~\rm{MHz}$, and $\Gamma_{20}=1~\rm{kHz}$.
  }
  \label{fig3}
\end{figure*}

\begin{figure}[htbp]
  \centering
  \includegraphics[width=0.4\textwidth]{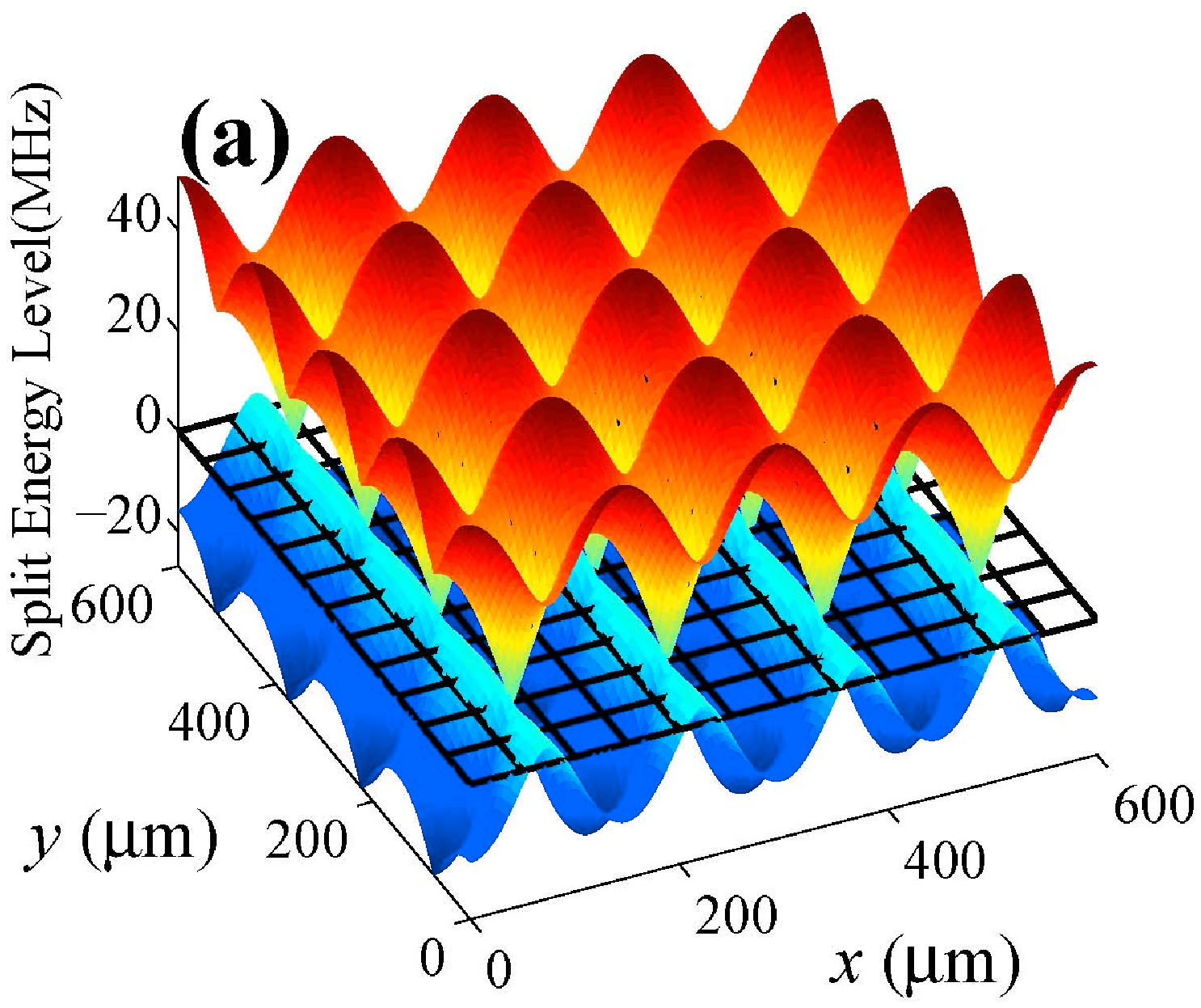}
  \includegraphics[width=0.4\textwidth]{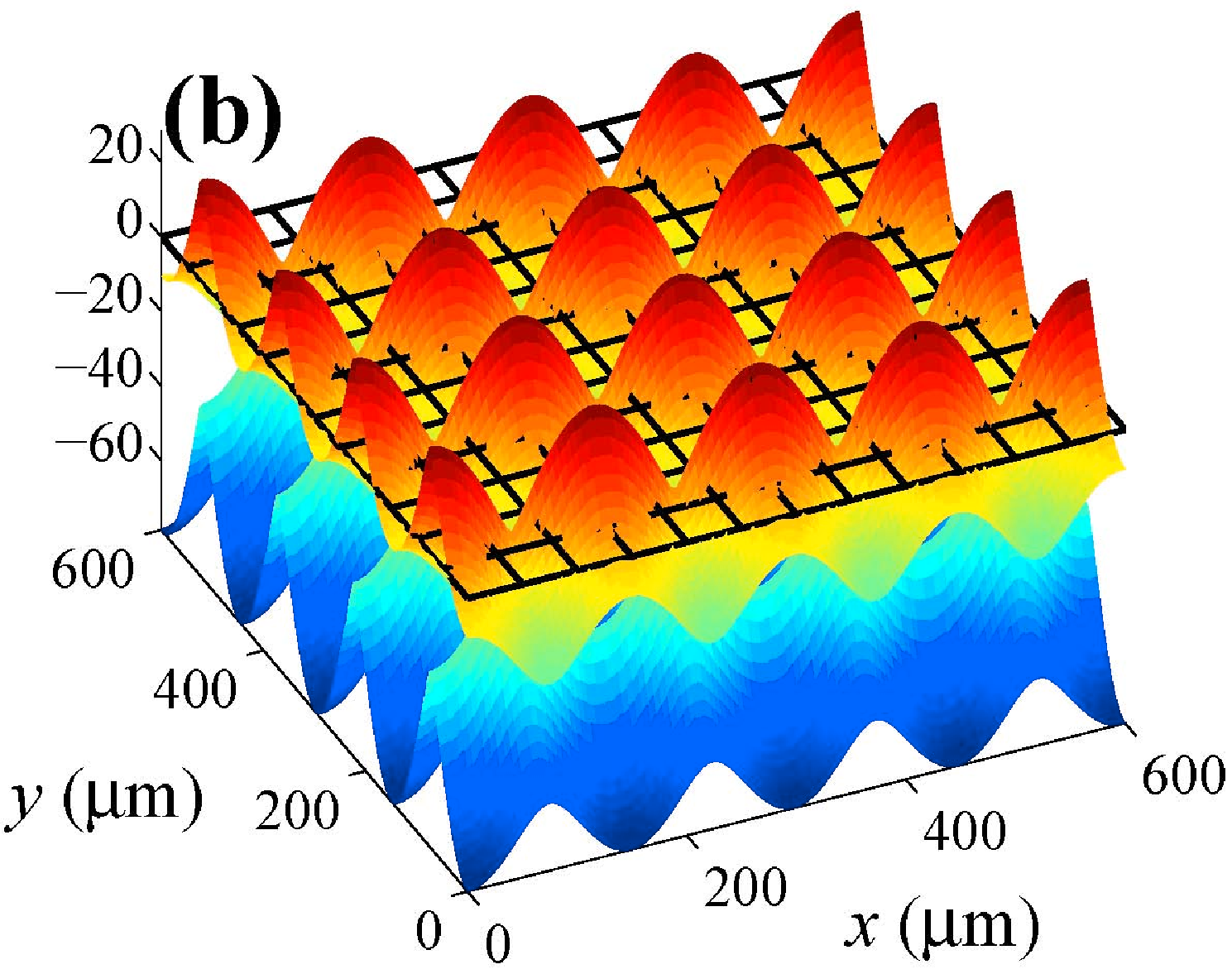}
  \caption{(a) The surfaces represent $\left| {G_{3t} \left( y \right) + G_{2t} \left( x \right) + } \right\rangle $
  (top), $|1\rangle$ (grid), and $\left| {G_{3t} \left( y \right) + G_{2t} \left( x \right) - } \right\rangle $ (bottom), respectively.
  (b) The surfaces are $\left| {G_{3t} \left( y \right) - G_{2t} \left( x \right) + } \right\rangle $
  (top), $|1\rangle$ (grid), and $\left| {G_{3t} \left( y \right) - G_{2t} \left( x \right) - } \right\rangle $ (bottom), respectively.
  $\Delta_2 = 2~{\rm{MHz}}$ for (a) and $-2~{\rm{MHz}}$ for (b).
  The values of the other parameters are $\Delta_3 = -3~{\rm{MHz}}$,
  $G_{20}  = G_{20}' = G_{30}  = G_{30}' = 15~{\rm{MHz}}$,
  $\Gamma_{10}  = 5~{\rm{MHz}}$,
  $\Gamma_{20}  = \Gamma_{30}  = 1~{\rm{kHz}}$,
  $\lambda_1  = 776~{\rm{nm}}$,
  $\lambda_2  = \lambda_3  = 780~{\rm{nm}}$,
  $\theta_1  = \theta_2  = 0.3^ \circ  $.
  }
  \label{fig4}
\end{figure}

Similar to the case for ${\bf E}_{F1}$,
we can also investigate the enhancement and suppression for ${\bf E}_{F2}$ and ${\bf E}_S$.
And their enhancement and suppression correspond to $\Delta_1  =  - \lambda_{|G_{3t}(y) \pm G_{2t}(x) \pm \rangle}$
and $\Delta_1  + \Delta_2  = 0$, respectively.
The periodic doubly-dressed FWM signals along $x$ corresponding to different conditions
are shown by the mesh plots in Figs. \ref{fig3}(a1)-(i1)
with insets being the dependence of the signal intensity on $\Delta_2$ at $z=0$ and $x=0$..
The corresponding split energy level schematics are shown in Figs. \ref{fig3}(a2)-(i2),
and the corresponding spatial periodic energy levels in Figs. \ref{fig3}(a3)-(i3).
Doubly-dressed SWM enhancement and suppression effects are similar to the results shown in Fig. \ref{fig3}.
In light of the SWM here is a two dimensional (2D) case,
we exhibit two cases of the second order splitting energy levels as shown in Figs. \ref{fig4}(a) and (b).

\section{Talbot Effect of MWM Signals}
\label{talbot}

In the perspective of Fourier optics,
the transfer function of a Fresnel diffraction system with $z$ as the propagation axis,
can be expressed as $H_F \left( \xi  \right) = \exp \left( {ik_{z} z} \right)\exp \left( { - i\pi \lambda z\xi ^2 } \right)$
in frequency domain \cite{goodman_2005},
where $\xi$ is the spatial frequency and ${k}_{z}$ is the projection of $\bf{k}$ along $z$.
The field of the MWM signal $g_0 \left( {x,y} \right) \propto \rho \left( {x,y} \right)$,
so it can be expanded into two-dimension (2D) Fourier series as
$
g_0 (x,y) = \sum_{m,n =  - \infty }^\infty  {c_{m,n} \exp } \left[i2\pi \left({nx}/{d_x} + {my}/{d_y} \right) \right],
$
and in frequency domain the equation above can be written as
\begin{equation}\label{eq11}
G_0 \left( {\xi ,\eta } \right) = \sum\limits_{m,n = - \infty }^\infty  c_{m,n} \delta \left(\xi  - \frac{n}{d_x} \right)  \cdot \delta \left(\eta  - \frac{m}{d_y}\right),
\end{equation}
where $c_{m,n}$ is the Fourier coefficient.
So, considering the Fresnel diffraction, the MWM signal at a $z$ distance is
\begin{equation}\label{eq12}
G\left( {\xi ,\eta } \right) = G_0 \left( {\xi ,\eta } \right)\exp \left( {ik_{z } z} \right)\exp \left[ { - i\pi \lambda_1 z\left( {\xi ^2  + \eta ^2 } \right)} \right].
\end{equation}
Plugging Eq. (\ref{eq11}) into Eq. (\ref{eq12}), we end up with
\begin{equation}\label{eq13}
G\left( {\xi ,\eta } \right) = \sum\limits_{m,n =  - \infty }^\infty  {c_{m,n} \delta \left( {\xi  - \frac{n}{{d_x }}} \right)}  \cdot \delta \left( {\eta  - \frac{m}{{d_y }}} \right)
\exp \left( {ik_{z} z} \right)\exp \left\{ { - i\pi \lambda_1 z\left[ {\left( {\frac{n}{{d_x }}} \right)^2  + \left( {\frac{m}{{d_y }}} \right)^2 } \right]} \right\}.
\end{equation}
For simplicity, we let $\lambda_2  = \lambda_3$ and $\theta_1  = \theta_2$,
thus the periods along $x$ and $y$ are the same, i.e., $d_x=d_y$.
Singly- and doubly-dressed FWM signals only concerns one SW formed by ${\bf k}_2$ and ${\bf k}_2'$,
and therefore we do not consider $y$ component.
In such a case, $\exp[- i\pi \lambda_1 z(n/d_x)^2] = 1$  if  $z=2qd_x^2/\lambda_1$ ($q=1,2,3,\cdots$),
so after inverse Fourier transformation, we get
\begin{equation}\label{eq14}
g\left( x \right) = g_0 (x)\exp \left( {ik_{z} z} \right).
\end{equation}
Because of $| {g\left( x \right)}|^2  = | {g_0 (x)} |^2$,
we can see the self-imaging of the MWM signals at $z = 2{{qd_x^2 }/{\lambda_1 }}$,
and $ {z_T } = z|_{q = 1}$ is the Talbot length.
It is worth mentioning that the images on the $z_T/N$ planes are fractional Talbot images \cite{chen_ol_2012},
where $N$ is an integer bigger than $1$.

\begin{figure}[htbp]
\centering
\includegraphics[width=0.8\textwidth]{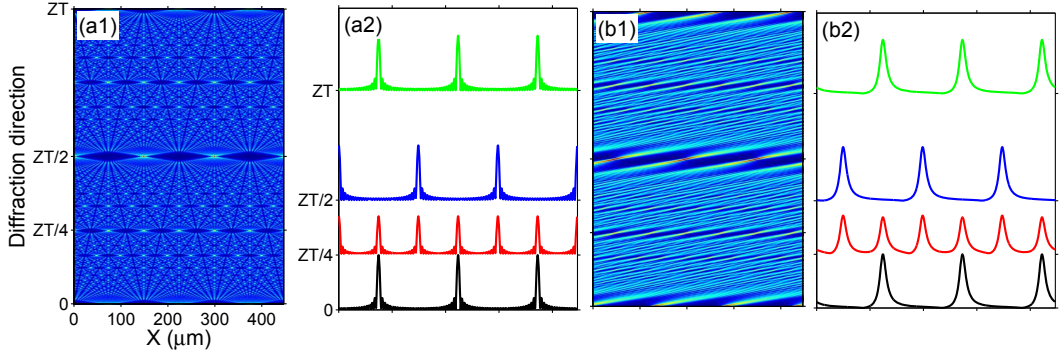}
\caption{
  (a1) The Talbot effect carpets for ${\bf E}_{F1}$ under suppression conditions with $\Delta_1 = 3~{\rm{MHz}}$ and $\Delta_2 = - 3~{\rm{MHz}}$.
  (b1) The Talbot effect carpets for ${\bf E}_{F1}$ under enhancement conditions with $\Delta_2 = - 3~{\rm{MHz}}$ and $ - \Delta_1=\Delta_2/2+ \sqrt{\Delta_2^2/4 + |G_{2t}(x)|^2} $.
  (a2) and (b2) The intensity profiles at $z=0$ (black curves),
  $z=z_T/4$ (red curves),
  $z=z_T/2$ (blue curves),
  and $z=z_T$  (green curves), respectively.
  The other parameters are
  $G_{20} = G_{20}' = 15~{\rm{MHz}}$,
  $\Gamma_{10}  = 5~{\rm{MHz}}$, and
  $\Gamma_{20}  = \Gamma_{30}  = 1~{\rm{kHz}}$.
}
\label{fig5}
\end{figure}

We firstly choose ${\bf E}_{F1}$ as the incidence to execute the simulation.
As shown in Fig. \ref{fig2}, the spatial profile of the incident signal varies along $\Delta_2$ with different $\Delta_1$,
and this variation leads to different diffraction process of FWM signal under different conditions.
Figs. \ref{fig5}(a1) and (b1) are the corresponding Talbot effect carpets
under suppression conditions and enhancement conditions, respectively.
Figs. \ref{fig5}(a2) and (b2) are the intensity profiles of the repeated images cutted
at certain fractional Talbot lengths.
From Fig. \ref{fig5}, we can not only clearly see the periodic singly-dressed FWM can reappear along  $z$,
but also see that the carpet stripes are oblique and such obliquity is more obvious under enhancement conditions.

\begin{figure}[htbp]
  \centering
  \includegraphics[width=0.8\textwidth]{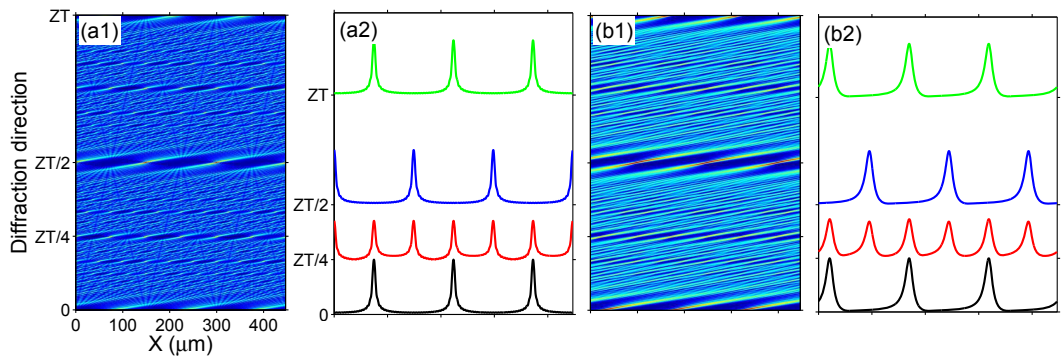}
  \caption{
  (a1) The Talbot effect carpets for ${\bf E}_{F2}$ under suppression conditions with
  $\Delta_1 = 18{\rm{~MHz}}$,
  $\Delta_2 = -18~{\rm{MHz}}$, and
  $\Delta_3 = 18{\rm{~MHz}}$.
  (b1) The Talbot effect carpets for ${\bf E}_{F2}$ under enhancement conditions with
  $\Delta_2 = -18~{\rm{MHz}}$,
  $\Delta_3 = 18{\rm{~MHz}}$, and $\Delta_1$ based on Eq. (\ref{eq10}).
  (a2) and (b2) The setup is as Figs. \ref{fig5}(a2) and (b2).
  The other parameters are
  $G_{20} = G_{20}' = 15~{\rm{MHz}}$,
  $G_{30}  = 2{\rm{~MHz}}$,
  $\Gamma_{10}  = 5~{\rm{MHz}}$, and
  $\Gamma_{20}  = \Gamma_{30}  = 1~{\rm{kHz}}$.
  }
  \label{fig6}
\end{figure}

For ${\bf E}_{F2}$, the results are shown in Fig. \ref{fig6}.
Clearly, we obtain the self-imaging of the incident ${\bf E}_{F2}$ at the Talbot plane again,
no matter it is under the suppression or the enhancement conditions.
In contrary to the case in Fig. \ref{fig5}(a1),
which is obtained under suppression conditions,
the Talbot effect shown in Fig. \ref{fig6}(a1) seems oblique.
Under enhancement conditions the carpet stripes are more obvious as shown in Fig. \ref{fig6}(b1),
but the obliquity is almost the same with that shown in Fig. \ref{fig6}(a1).
By comparing the Talbot carpets shown in Fig. \ref{fig6}(b1) and Fig. \ref{fig5}(b1),
we find the obliquity and the width of the stripes are almost unchanged.

\begin{figure}[htbp]
\centering
\includegraphics[height=2.5cm]{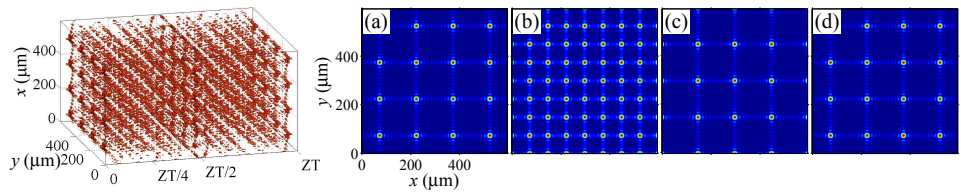}
  \caption{
  The iso-surface plot is the Talbot effect for ${\bf E}_{S}$ under the suppression condition.
  The four panels are the contour plots of the Talbot effect at $z=0$ (a),
  $z=z_T/4$ (b), $z=z_T/2$ (c), and $z=z_T$ (d), respectively.
  The parameters are
  $\Delta_1 = 18~{\rm{MHz}}$,
  $\Delta_2 =- 18~{\rm{MHz}}$,
  $\Delta_3 = 18~{\rm{MHz}}$,
  $G_{20} = G_{20}' = G_{30} = G_{30}' = 15~{\rm{MHz}}$,
  $\Gamma_{10}  = 5~{\rm{MHz}}$,
  $\Gamma_{20}  = \Gamma_{30}  = 1~{\rm{kHz}}$,
}
\label{fig7}
\end{figure}

\begin{figure}[htbp]
  \centering
  \includegraphics[height=2.5cm]{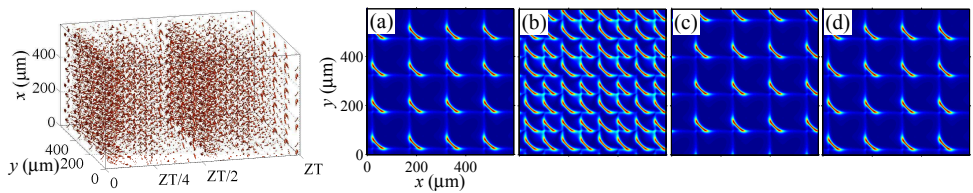}
  \caption{
  The Talbot effect for ${\bf E}_{S}$ under the enhancement condition.
  The setup is as the Fig. \ref{fig7}, but with
  $\Delta_1$ according to Eq. (\ref{eq9}).
  }
\label{fig8}
\end{figure}

Last but not the least, we discuss the Talbot effect from ${\bf E}_{S}$.
We consider two orthogonal SWs from two couple of dressing fields  ${\bf E}_2$, ${\bf E}_2'$  and ${\bf E}_3$, ${\bf E}_3'$  simultaneously,
to form a 2D lattice,
which is periodic both along $x$ and $y$ as shown in Fig. \ref{fig1}(d),
thus a 2D SWM signal will be excited.
In Figs. \ref{fig7} and \ref{fig8}, we first give the iso-surface plots of the Talbot effect of the 2D SWM signal
under suppression conditions and enhancement conditions, respectively.
And then we choose four intensity plots at certain places during propagation to show the details more clearly.
We can see that at the Talbot length as shown in Figs. \ref{fig7}(d) and \ref{fig8}(d),
the 2D SWM signals are reproduced.
At half the Talbot length as shown in Figs. \ref{fig7}(c) and \ref{fig8}(c),
the self-images shifted half period both along $x$ and $y$.
At one quarter of the Talbot length, fractional self-images can be seen as shown in Figs. \ref{fig7}(b) and \ref{fig8}(b),
in which the images are twice as many as those in Figs. \ref{fig7}(a) and \ref{fig8}(a).
Under enhancement conditions shown in Fig. \ref{fig8},
the images are more clear than those under suppression conditions as shown in Fig. \ref{fig7}.
Here, we want to point out the reason that the periods of fractional Talbot effects shown in
Figs. \ref{fig5}(b2), \ref{fig6}(a2) and (b2), and \ref{fig8}(b) seemingly not the half period at $z=0$,
is due to the inertial weakness of the numerical simulations.
The Talbot effect changes dramatically during propagation, especially around small fractional Talbot lengths.
Even a tiny deviation from the exact $z_T/4$ will bring seemingly imperfect quarter Talbot effect.
Calculations with very high resolution will help us approach perfect fractional Talbot effects,
but beyond our computer's memory.

\section{Conclusion}
In conclusion, we have studied the Talbot effects with completely controllable MWM signals.
We have obtained the spatially periodic FWM and SWM signals by the interference between two dressing fields in a reverse Y-type atomic level system.
The intensities of the MWM signals can be effectively controlled via enhancement conditions and suppression conditions
derived from dressing effect.
Talbot effect from singly- as well as doubly-dressed FWM is observed in the numerical experiment.
Different from the case of FWM, for SWM, we construct a 2D lattice,
from which the SWM is 2D modulated, to investigate the 2D Talbot effect.
We find that the numerical simulations agree with the theoretical predictions very well.
Our scheme is more advantageous in the controllability, compared to the previous studies,
mainly attributed to the modulation of the MWM signal by the dressing effect.
Our findings not only enrich the understanding of the MWM theory,
but also offer a different method to investigate the Talbot effect.

\section*{Acknowledgement}
This work was supported in part by
the 973 Program 2012CB921804; by the NNSFC under Grants 10974151, 61078002, 61078020,
11104214, 61108017, and 11104216; by the NCET under Grant 08-0431; by the RFDP under Grants
20110201110006, 20110201120005, and 20100201120031; and by the FRFCU under Grants
2011JDHZ07, XJJ2011083, XJJ2011084, XJJ20100151, XJJ20100100, and XJJ2012080.

\end{document}